\documentclass[twocolumn,english,aps,prl,groupedaddress,superscriptaddress,showpacs]{revtex4}
\usepackage[latin9]{inputenc}
\usepackage{amsmath}
\usepackage{amssymb}
\usepackage{graphicx}
\usepackage{hyperref}
\hypersetup{colorlinks,citecolor=blue,linkcolor=blue,urlcolor=blue}
\usepackage{amsbsy}\usepackage{units}\usepackage{xcolor}\usepackage{amsfonts}\usepackage{mathrsfs}\usepackage{bm}\usepackage{babel}

\newcommand{\subtext}[1]{\ensuremath{_{\mbox{\scriptsize #1}}}}
\DeclareRobustCommand{\bafeTMasx}[3]{Ba(Fe$_{#3}$#1$_{#2}$)$_{2}$As$_{2}$}
\newcommand{\cafeas}{CaFe$_2$As$_2$}
\newcommand{\bafeas}{BaFe$_2$As$_2$}
\newcommand{\srfeas}{SrFe$_2$As$_2$}
\DeclareRobustCommand{\bafecoasx}[2]{\bafeTMasx{Co}{#1}{#2}}
\DeclareRobustCommand{\bafecoas}{\bafecoasx{x}{1-x}}

\newcommand{\bfQ}[1]{\ensuremath{\mathbf{Q}\subtext{#1}}}
\newcommand{\bfq}[1]{\ensuremath{\mathbf{q}\subtext{#1}}}

\newcommand{\chiim}[1]{\ensuremath{\chi''\subtext{#1}\!\!\left(\bfQ{},E\right)}}

\newcommand{\Eq}[1]{Eq.~(\ref{#1})}
\newcommand{\Eqs}[1]{Eqs.~(\ref{#1})}
\newcommand{\plaineq}[1]{(\ref{#1})}

\newcommand{\Fig}[1]{Fig.~\ref{#1}}
\newcommand{\Figs}[1]{Figs.~\ref{#1}}

\begin{document}

\title{Crossover from spin-waves to diffusive spin excitations in underdoped \bafecoas{}}
\author{G.~S.~Tucker}
\affiliation{Ames Laboratory, U.S. Department of Energy, Ames, IA, 50011, USA}
\affiliation{Department of Physics and Astronomy, Iowa State University, Ames, IA, 50011, USA}
\author{R.~M.~Fernandes}
\affiliation{School of Physics and Astronomy, University of Minnesota, Minneapolis, MN, 55455, USA}
\author{D.~K.~Pratt}
\author{A.~Thaler}
\author{N.~Ni}
\affiliation{Ames Laboratory, U.S. Department of Energy, Ames, IA, 50011, USA}
\affiliation{Department of Physics and Astronomy, Iowa State University, Ames, IA, 50011, USA}
\author{K.~Marty}
\author{A.~D.~Christianson}
\author{M.~D.~Lumsden}
\affiliation{Quantum Condensed Matter Division, Oak Ridge National Laboratory, Oak Ridge, TN, 37831, USA}
\author{B.~C.~Sales}
\author{A.~S.~Sefat}
\affiliation{Materials Science and Technology Division, Oak Ridge National Laboratory, Oak Ridge, TN, 37831, USA}
\author{S.~L.~Bud'ko}
\author{P.~C.~Canfield}
\author{A.~Kreyssig}
\author{A.~I.~Goldman}
\author{R.~J.~McQueeney}
\affiliation{Ames Laboratory, U.S. Department of Energy, Ames, IA, 50011, USA}
\affiliation{Department of Physics and Astronomy, Iowa State University, Ames, IA, 50011, USA}

\begin{abstract}
Using inelastic neutron scattering, we show that the onset of superconductivity
in underdoped Ba(Fe$_{1-x}$Co$_{x}$)$_{2}$As$_{2}$ coincides with
a crossover from well-defined spin waves to overdamped and diffusive
spin excitations. This crossover occurs despite the presence of long-range
stripe antiferromagnetic order for samples in a compositional range
from $x=0.04-0.055$, and is a consequence of the shrinking spin-density
wave gap and a corresponding increase in the particle-hole (Landau)
damping. The latter effect is captured by a simple itinerant model
relating Co doping to changes in the hot spots of the Fermi surface.
We argue that the overdamped spin fluctuations provide a pairing mechanism
for superconductivity in these materials.
\end{abstract}


\date{\today}

\maketitle
The key to many unconventional superconductors lies in their proximity to an ordered antiferromagnetic (AFM) phase \cite{Pines92,Chubukov03,Scalapino12}.
As a ground state that competes with superconductivity (SC), the suppression of AFM order (by chemical tuning or applied pressure) is required for the SC state to appear.
However, the vestiges of AFM order that remain as correlated spin fluctuations have been proposed to provide the glue that pairs electrons in the SC state \cite{magnetic}.
It is, therefore, very important to understand how the spin excitations evolve from collective spin waves in the AFM ordered state 
to the overdamped correlated spin fluctuations characteristic of the SC state.
In the iron pnictide \bafecoas{}, the suppression of AFM ordering upon Co substitution of a few percent allows a SC ground state to appear \cite{Ni08} in the presence of substantial spin fluctuations at the AFM wavevector, $\bfQ{AFM}$. 
Unlike some unconventional superconductors, e.g., Ba$_{1-x}$K$_x$Fe$_2$As$_2$, the competing AFM ordered and SC states actually coexist microscopically in a limited compositional range from $x\approx0.04-0.06$, the so-called underdoped compositions \cite{FernandesPRB10}.
This allows one to investigate how the normal state spin fluctuations provide the conditions for SC to emerge, even in the presence of weak AFM order.

Given the important connection between superconductivity and magnetism, extensive studies of the magnetic dynamics have been performed in these compounds as a function of composition.
The magnetic dynamics of electron doped compounds, \bafeTMasx{$M$}{x}{1-x} ($M=$ Co, Ni) have been studied in some detail by inelastic neutron scattering (INS) \cite{Lester10,Li10,Harriger11,Dai12,Tucker12,Liu12}.
These investigations found that the high-energy spin dynamics ($E>50$ meV) are relatively insensitive to electron doping whereas the low-energy spin dynamics show a strong dependence on electron doping.
Deep within the AFM ordered state of the parent \bafeas{} compound (N\'eel transition temperature, $T\subtext{N}=136$ K), the low-energy spin dynamics are dominated by a large spin gap $\Delta\approx10$ meV that characterizes the ordered AFM state \cite{Sato09}. 
Above the spin gap, very steep spin wave excitations propagate within the Fe layer, while much lower spin wave velocities are found for modes propagating perpendicular to the layers, indicative of quasi-two-dimensional magnetism.
In \bafeas{} \cite{Ewings08}, as well as \cafeas{} \cite{McQueeney08} and \srfeas{} \cite{Zhao08}, the low-energy spin waves have very long lifetimes (no substantial energy-dependent damping). 
The large spin gap and small damping of the collective spin wave modes highlight the robust AFM state of the parent compounds. 
Within an itinerant spin-density wave picture for the AFM order in the iron pnictides, such behavior indicates that the electronic spin-density wave (SDW) gap is large, estimated to be $\Delta_{\mathrm{SDW}}>50$ meV via optical conductivity measurements \cite{Hu08}, thereby gapping out particle-hole (Landau) damping mechanisms. 
Note that while the spin gap $\Delta$ is related to anisotropies in spin space (such as single-ion anisotropy), the SDW gap $\Delta_{\mathrm{SDW}}$ is proportional to the magnetization and, therefore, to the energy gain in the magnetically ordered state.

At the opposite extreme, those compositions without long-range AFM order ($x>0.06$ for Co substitutions) display low-energy spin excitations that are diffusive (overdamped) in nature, and typical of systems close to a critical point \cite{Inosov10,Li10}. 
The low-energy spin fluctuations are still centered at $\textbf{Q}_{\mathrm{AFM}}$, but appear gapless and are characterized by a finite spin-spin correlation length ($\xi$) and relaxational energy scale ($\Gamma$) related to the Landau damping. 
The presence of substantial magnetic spectral weight at low energies (as obtained from a gapless spectrum with $\Gamma\sim k\subtext{B}T\subtext{c}$, with superconducting transition temperature $T\subtext{c}$) is considered an important ingredient for magnetically-mediated SC \cite{Chubukov03}.

Here, we study the evolution of the normal state spin dynamics between these two extremes.
Most of the compositions are underdoped, possessing both weak AFM ordering and superconductivity at low temperatures (i.e. small SDW and SC gaps). 
In the normal state of the underdoped compounds, we find clear signatures of diffusive behavior in the low-energy spin dynamics (spatial disorder and a gapless spectrum with overdamped dynamics) despite the AFM ordering. 
The crossover of the spin dynamics is associated both with the collapse of the spin-density wave gap, see Ref.~\onlinecite{Sachdev95}, and the subsequent development of strong Landau (particle-hole) damping. 
This crossover coincides with the appearance of SC in underdoped samples.

The INS measurements were carried out on the HB3 spectrometer at the High Flux Isotope Reactor at the Oak Ridge National Laboratory. 
Samples were grown and characterized as outlined in Ref.~\onlinecite{Ni08} and were mounted in the $[1,1,0]$--$[0,0,1]$ scattering plane.
We define 
$\mathbf{Q}$ = $\frac{2\pi}{a}H\hat{\mathrm{i}}+\frac{2\pi}{a}K\hat{\mathrm{j}}+\frac{2\pi}{c}L\hat{\mathrm{k}}$ = $(H,K,L)$ 
in reciprocal lattice units as referenced to the tetragonal $I4/mmm$ unit cell. 
Details about the instrumental configuration and resolution are given elsewhere.

\begin{figure}
\includegraphics{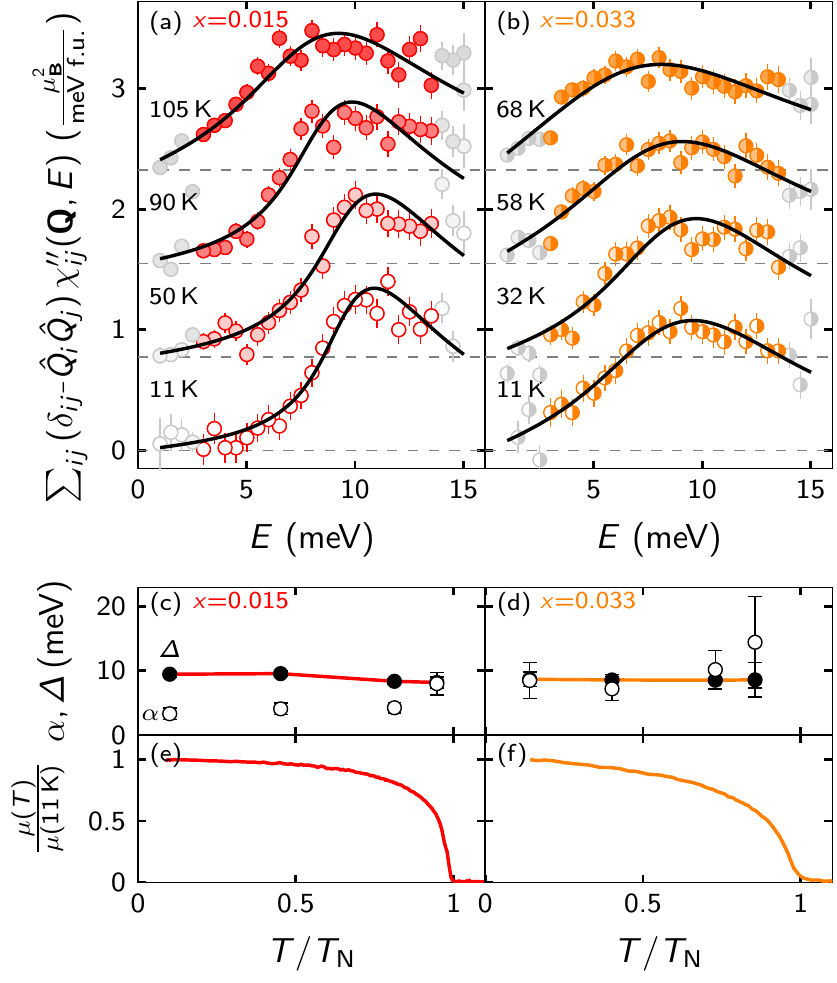} 
\caption{
Temperature dependence of INS data for \bafecoas{} with $x=0.015$ [(a,\,c,\,e)] and $x=0.033$ [(b,\,d,\,f)] plus fits to the spinwave model.
(a,\,b) Energy scans at $\bfQ{AFM}=(0.5,0.5,1)$ performed at the indicated temperatures are offset vertically.
(c,\,d) Reduced temperature dependence of spinwave model parameters $\alpha$ (open symbols) and $\Delta$ (filled symbols).
(e,\,f) Reduced temperature dependence of the ordered magnetic moment, $\mu$, normalized to its low-temperature value.
Light gray symbols in panels (a) and (b) represent measured intensity which was excluded from fitting due to concerns with the validity of background estimates at those points.
\label{fig:tempdep} }
\end{figure}

Figure \ref{fig:tempdep} shows the $\bfQ{AFM}=(0.5,0.5,1)$ spectrum of \bafecoas{} at several different temperatures for lightly doped and nonsuperconducting $x=0.015$ and $0.033$.
The spectra are dominated by a large spin gap at $\sim$ 10 meV for both compositions. 
These data can be fit to a damped spin wave form for $\chiim{}$, 
\begin{equation}
\chiim{s} \propto \frac{E}{{\left(\Delta^2+c^2q^2-E^2\right)^2+\alpha^2E^2}},
\label{eqn1}
\end{equation}
where $\Delta$ is a spin gap, $c$ is the spin wave velocity, $\alpha$ is a damping rate, and $\bfq{}\equiv\bfQ{}-\bfQ{AFM}$ is the reduced momentum transfer. 
The full anisotropic form for the damped spin wave susceptibility is given elsewhere.
For $x=0.015$ at $11$ K, $\alpha=3.6(4)$ meV is small in comparison to other energy scales and, in principle, can arise from a combination of different damping processes (such as Landau damping for energy scales larger than the SDW gap, or magnon-magnon interactions). 
The fit to the $x=0.015$ $11$ K data shows a large spin gap $\Delta=9.73(14)$ meV characteristic of the parent AFM ordered state.

The solid lines in Fig.~\ref{fig:tempdep} (a) and (b) represent independent fits to the damped spin wave model where the gap and damping rate are allowed to vary freely. 
The magnitude of the spin gap is determined to be nearly constant with temperature up to our closest approach of $T/T\subtext{N}=0.95$ where the ordered magnetic moment $\mu(T)/\mu(11\text{ K})\approx0.5$. 
Similar to the results described for NaFeAs \cite{Park12}, \bafeas{} \cite{Park12}, and LaFeAsO \cite{Ramazanoglu13}, we find that the spin gap energy, $\Delta$, is roughly 9 meV in the ordered state, regardless of size of the ordered moment or the concentration, $x$, and the dynamics become overdamped as $T\subtext{N}$ is approached.

\begin{figure}
\includegraphics{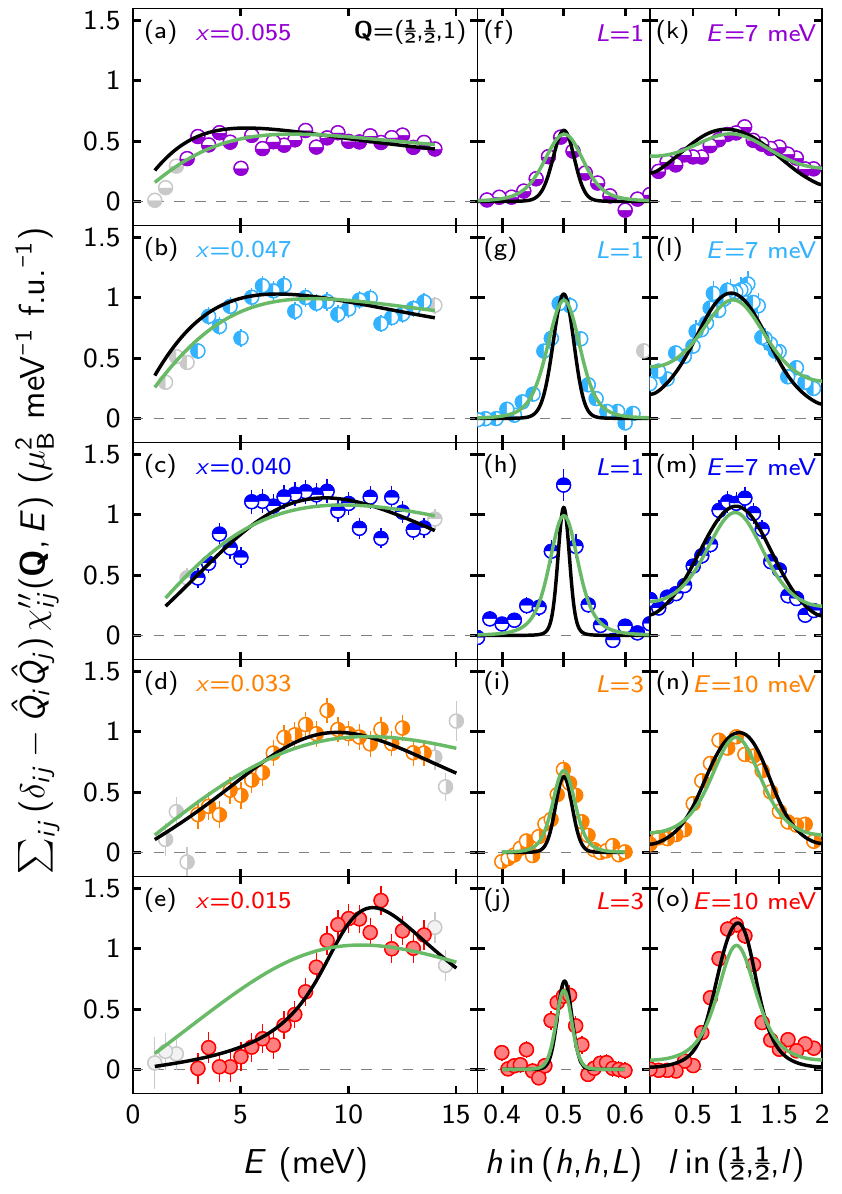} 
\caption{
Background subtracted INS intensity of \bafecoas{} corrected for the Bose thermal population factor and the Fe$^{2+}$ single-ion magnetic form factor 
plus best fit lines to the diffusive (light green lines) and the damped spin-wave (black lines) models. 
(a-e) Constant-\bfQ{} energy scans at $\bfQ{AFM}=(0.5,0.5,1)$ for five compositions.
(f-h) Constant-$E$ scans in the $[h,h,0]$-direction across $\bfQ{AFM}=(0.5,\,0.5,\,1)$ at $E=7$ meV. 
(i-j) Constant-$E$ scans in the $[h,h,0]$-direction across $\bfQ{AFM}=(0.5,\,0.5,\,3)$ at $E=10$ meV. 
(k-m) Constant-$E$ scans in the $[0,0,l]$-direction, perpendicular to the Fe layer, across $\bfQ{AFM}=(0.5,0.5,1)$ at $E=7$ meV. 
(n-o) Constant-$E$ scans in the $[0,0,l]$-direction across $\bfQ{AFM}=(0.5,\,0.5,\,1)$ at $E=10$ meV. 
Light gray symbols represent measured intensity which was excluded from fitting due to concerns with the validity of background estimates at those points.
\label{fig:scans} }
\end{figure}

Figure~\ref{fig:scans} shows a series of representative low-energy INS scans taken in the AFM ordered and normal state ($T\subtext{c}<T<T\subtext{N}$) for each composition. 
Upon increased Co substitution, the spin gap appears to gradually close [\Figs{fig:scans}(a)-(e)] and is completely absent at $x=0.055$. 
One can also observe a gradual reciprocal space broadening of the longitudinal cut [\Figs{fig:scans}(f)-(j)] with increasing Co composition. 
Finally, the modulations along $(1/2,1/2,l)$ [\Figs{fig:scans}(k)-(o)] are reduced, signaling a gradual evolution to two-dimensional spin dynamics. 
Within the damped spin wave model of \Eq{eqn1}, the data at all compositions 
have been successfully 
fit by assuming that: in accordance with our temperature-dependent results, the spin gap remains constant; the damping increases dramatically with $x$; and both the in-plane and inter-plane spin wave velocities are reduced with $x$. 
However, it is clear from high-energy INS investigations that the in-plane spin velocities are independent of composition (see the discussion in Ref.~\onlinecite{Tucker12}) and constraining the in-plane velocity to this value leads to poorer and poorer agreement of the low-energy data with the damped spin wave model (as shown by the black lines in \Fig{fig:scans}).

One major assumption of our data analysis using the spin wave model is that the spin gap is independent of composition. 
If the spin gap is due to single-ion anisotropy, then its magnitude should be proportional to some power of $\mu$ \cite{Fishman98}. 
Data fitting in which the spin gap was allowed to freely vary resulted in an \emph{increase} of the gap with composition, and fits in which the spin gap was constrained to be proportional to $\mu$ gave worse results.

The increased reciprocal space broadening suggests that another length scale must be introduced for low-energy magnetic fluctuations, such as a spin-spin correlation length. 
Considering also the gapless form of the magnetic excitations, the data at higher compositions resemble the diffusive response 
that has been used to describe the optimal and overdoped samples.\cite{Li10,Matan10} 
This diffusive response has the form 
\begin{align}
\chiim{d} &\propto \frac{E}{a^4\left(q^2+\xi^{-2}\right)^2+\gamma^2E^2} \\
          &=\frac{E}{\Gamma^2\left(1+q^2\xi^{2}\right)^2+E^2 },
\label{eqn2}
\end{align}
where $\xi$ is the spin-spin correlation length, $a$ is the tetragonal lattice constant, and $\gamma$ is the Landau damping coefficient. 
One can also define $\Gamma\equiv a^{2}/\gamma\xi^{2}$ as the spin relaxation rate. 
Fits to the diffusive form for $\chiim{}$ are shown as light green lines in \Fig{fig:scans}. 
While the diffusive form does a poor job at the lowest compositions where the spin gap is sharp, it works exceptionally well at the higher compositions where the spectrum appears gapless and the increased reciprocal space broadening for longitudinal scans shown in \Fig{fig:scans}(f)-(j) is captured by a smaller correlation length.

\begin{figure}
\includegraphics{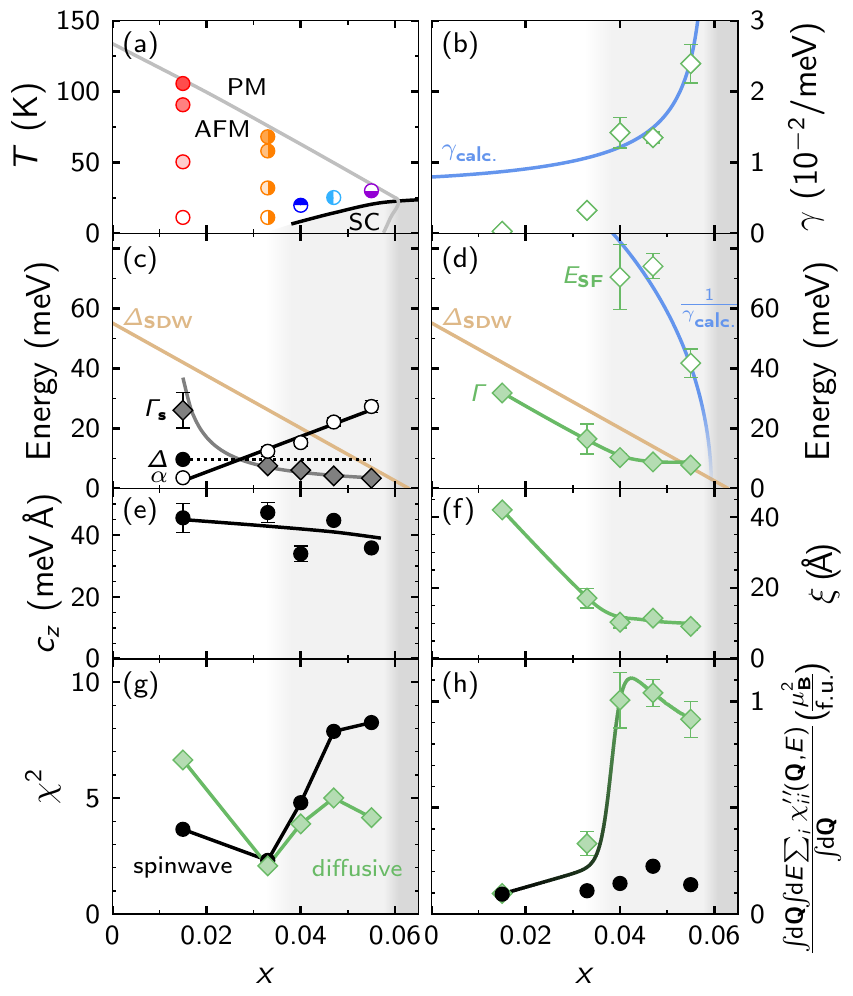}
\caption{
(a) Phase diagram of \bafecoas{} showing regions of AFM, SC, and their coexistence; colored symbols show the locations in phase-space of the measurements performed in this study. 
(b-f) Select model parameters as a function of composition for the diffusive (green diamonds) and spin wave (black circles) models. 
All data points shown in (b-h) were determined at the lowest temperature indicated in (a), and the lightly shaded background indicates compositions which exhibit SC at low temperature. 
Spin-wave model: (c)  spin gap $\Delta$ (filled), damping $\alpha$ (open), and $\Gamma\subtext{s}=\Delta^2/\alpha$ (diamonds); (e) Inter-plane spin wave velocity. 
Diffusive model: (b) Landau damping $\gamma$ and the corresponding theoretical prediction; 
(d) Spin relaxation characteristic energy $\Gamma$ (filled) and the effective magnetic energy $E\subtext{SF}=1/\gamma$ (open);
(f) correlation length $\xi$.   
In (c-d) an estimate for the SDW gap -- derived from $\mu(x)$ from Ref.~\onlinecite{FernandesPRB10} and $\Delta\subtext{SDW}(x=0)$ from Ref.~\onlinecite{Hu08} -- is also shown (tan line).
(g) Residual $\chi^{2}$ for each fit model. 
(h) Spectral weight of the $(0.5,0.5,1)$ excitation. 
The spectral weight is the $\bfQ{}$-averaged energy integration of the trace of the imaginary component of the magnetic susceptibility tensor. 
The averaging-range in $\bfQ{}$ here is $0\!<\!H\!<\!1$, $0\!<\!K\!<\!1$, $0\!<\!L\!<\!2$; the energy integration is over the range $0\!<\!E\!<\!35$ meV. 
All errorbars represent the combined errors for all function parameters. 
The solid green and black lines in (c-h) are guides to the eye. 
\label{fig:params} 
}
\end{figure}

Figure~\ref{fig:params} shows the locations of our measurements in a phase-space diagram, the fitting parameters for both the damped spin wave and diffusive models in \Eqs{eqn1} and \plaineq{eqn2}, a $\chi^{2}$ measure of the goodness-of-fit for the constant-\bfQ{} and constant-$E$ scans presented in Fig.~\ref{fig:scans} for these two models, and the composition-dependence of the low-energy spectral weight. 
For $x=0.015$, the damped spin wave model is the best and $\alpha/\Delta=0.37(8)$ is consistent with underdamped dynamics.
For intermediate composition, $x=0.033$, both models are of comparable quality. 
As shown in \Fig{fig:params}(c), within the damped spin wave model $\alpha/\Delta>1$ and the dynamics have become overdamped causing the spin gap to disappear.
In the limit where $\alpha/\Delta\gg1$, the overdamped spin wave model also takes on a relaxational form with $\Gamma\subtext{s}=\Delta^2/\alpha$; as shown in \Figs{fig:params}(c) and (d) $\Gamma\subtext{s}$, $\Gamma$, and the effective magnetic energy, $E\subtext{SF}=1/\gamma$, decrease as the critical concentration 
for which the AFM order is fully suppressed 
is approached, as indicated by vanishing $\Delta\subtext{SDW}$.
As seen in \Figs{fig:scans}(k-o) the excitation becomes increasingly two-dimensional with $x$ as captured by the damped spin-wave model parameter $c\subtext{z}$, \Fig{fig:params}(e).
For $x=0.040$, $0.047$, and $0.055$, the diffusive model becomes the better fit, as the smaller correlation length [\Fig{fig:params}(f)] is able to capture the reciprocal space broadening of the in-plane spin fluctuations.
A comparison of the residual for each fit model in \Fig{fig:params}(g) clearly shows the crossover from spin-wave- to diffusive-like excitations.
In \Fig{fig:params}(h), a sharp increase in the low-energy spectral weight ($<35$ meV) coincides with the appearance of SC.

From \Fig{fig:params}, regardless of the model used to fit
our data, it is clear that upon approaching the optimally-doped composition,
damping becomes stronger, the spin fluctuations acquire a more two-dimensional
character, and the energy scale associated with these fluctuations
($\Gamma$ or $\Gamma\subtext{s}$) become smaller. These features,
as well as the crossover from spin-wave to diffusive excitations,
are consistent with a suppression of the SDW gap $\Delta_{\mathrm{SDW}}$
upon doping. In \Figs{fig:params}(c-d), we show the experimentally
determined suppression of $\Delta_{\mathrm{SDW}}$ obtained by combining
the doping evolution of the zero-temperature ordered magnetic moment, $\mu(x)$, from
Ref.~\onlinecite{FernandesPRB10} with the optical conductivity derived value of $\Delta\subtext{SDW}(x=0)$ from
Ref.~\onlinecite{Hu08}, using the fact that $\Delta\subtext{SDW}\propto \mu$
\cite{FernandesPRB10,Fernandes_Schmalian,Vorontsov10,Eremin10}.

Based on this information, we can conclude that the presence of sub-gap
spectral weight which appears with either an increase in temperature
or Co composition is driven entirely by damping. For the temperature-driven
transition, we find an increase of damping close to $T\subtext{N}$.
Given the similarities between the spin fluctuations above and below $T_{N}$ near optimal
doping and the smallness of the spin-wave gap $\Delta_{\mathrm{SDW}}$
in this regime [see \Fig{fig:params}(c)], we compare the fitted
damping rate $\gamma$ with the calculated Landau damping $\gamma_{\mathrm{calc}}$
due to the decay of spin excitations into particle-hole pairs near
the Fermi level in \Fig{fig:params}(b). Using a simplified two-band
model, which was previously shown to successfully capture the coexistence
of SC and AFM \cite{Fernandes_Schmalian,Vorontsov10,Eremin10}, the
Landau damping is given by $\gamma_{\mathrm{calc}}^{-1}\propto\left|\mathbf{v}_{e}\times\mathbf{v}_{h}\right|$
\cite{Sachdev_book}, where $\mathbf{v}_{e}$ and $\mathbf{v}_{h}$
are respectively the Fermi velocities of the electron and hole pockets
at the hot spots (i.e. points connected by the AFM ordering vector
$\bfQ{AFM}$). Upon Co substitution, electrons are introduced into the system,
making the hole pocket shrink and the electron pocket expand. As revealed
by ARPES \cite{Liu10}, this moves the hot spots, making their Fermi
velocities become nearly parallel around optimal doping. As a result,
$\gamma_{\mathrm{calc}}^{-1}\rightarrow0$, as seen experimentally.
Note that $\gamma_{\mathrm{calc}}$ describes well the data only in
compositions near optimal doping, indicating that in slightly-doped
compositions the damping comes from another mechanism, such as magnon-magnon
interactions.

In summary, we have shown that in \bafecoas{} the low-energy spin dynamics, which are most strongly tied to excitations in close proximity to the Fermi surface, display a crossover from gapped spin waves to a regime of strong damping and short correlation length, even though weak AFM order persists. 
The appearance of strong Landau damping near $x=0.03$ -- $0.04$ coincides with the appearance of superconductivity, suggesting that the corresponding increase of low energy spectral weight below the spin gap is a key ingredient for superconductivity to develop. 
In theories where pairing is mediated by spin fluctuations, their energy scale ($E\subtext{SF}$) is usually positively correlated to $T\subtext{c}$ \cite{Pines92,Chubukov03,Scalapino12}.
We instead observe that $E\subtext{SF}$ decreases with increasing $x$ (and $T\subtext{c}$), see \Fig{fig:params}(d).
To avoid the apparent contradiction one must also consider that AFM and SC order compete \cite{Pratt09Christianson09} thereby effectively decreasing $T\subtext{c}$ for underdoped samples (and eliminating SC for the parent compound).
Indeed, when the long-range AFM order is suppresed by pressure, $T\subtext{c}$ for lower $x$ samples is enhanced beyond that for optimal doping\cite{Colombier10Ni10}.
Therefore, $T\subtext{c}$ and $E\subtext{SF}$ are infact positively correlated.
Iron pnictide compositions on either side of the SC region --- such as \bafecoas{} with $x=0.015$ or $x=0.14$ \cite{Sato11}, or \bafeTMasx{Ni}{0.15}{0.85} \cite{Wang13} --- lack overdamped spin fluctuations, in contrast to the underdoped SC compositions presented here;
this provides further evidence that overdamped spin fluctuations are a necessary component in the paring mechanism for superconductivity in the iron pnictides.

\begin{acknowledgments}
The authors would like to thank D. C. Johnston and A. Kaminski for useful discussions. 
Work at the Ames Laboratory was supported by the Department of Energy -- Basic Energy Sciences, Division of Materials Sciences and Engineering under Contract No. DE-AC02-07CH11358.
Part of the research conducted at ORNL's High Flux Isotope Reactor was sponsored by the Scientific User Facilities Division, Office of Basic Energy Sciences, U.S. Department of Energy.
Some work at Oak Ridge (BCS, ASS) was supported by the Department of Energy, Basic Energy Sciences, Materials Sciences and Engineering Division.
\end{acknowledgments}

\end{document}